\documentclass[preprint, aps,amsmath,amssymb,floats,12pt]{revtex4}
\usepackage{graphicx}
\usepackage{amsmath}

\begin{document}
\bibliographystyle{unsrt}
\providecommand{\bysame}{\leavevmode\hbox to3em{\hrulefill}\thinspace}
\providecommand{\MR}{\relax\ifhmode\unskip\space\fi MR }
% \MRhref is called by the amsart/book/proc definition of \MR.
\providecommand{\MRhref}[2]{%
  \href{http://www.ams.org/mathscinet-getitem?mr=#1}{#2}
}
\providecommand{\href}[2]{#2}
\pagestyle{empty}
\bibliographystyle{amsplain}

\title{Applying the Maximum Entropy Technique to the Gaussian Dispersion Plume Model}

\author{J.A. Secrest$^1$, J.M. Conroy$^2$ and  H.G. Miller$^{2}$ }

\affiliation{$^1$Department of Physics and Astronomy, Georgia Southern University, Armstrong Campus, Savannah, Georgia, USA}

\affiliation{$^2$Department of Physics, State University of New York at Fredonia, Fredonia, New York,
USA}

\date{\today}

\begin{abstract} The Maximum Entropy (MaxEnt) technique is applied to the derivation of the Gaussian Dispersion Plume Model  as well as to  more complex transport phenomena such as the one-dimensional advection equation, the one-dimensional diffusion equation, the one dimensional advection-diffusion equation, and finally to the multi-dimensional advection-diffusion equation.  Further application is discussed.

\end{abstract}

\maketitle

%%%%%%%%%%%%%%%%%%%%%%%%%%%%%%% Begin Manuscript
\section{Introduction}
The transport of contaminates due to advective and diffusive processes in the enviormental media from the emission of pollutant sources is a concern to many.  One of the most ubiquitous air pollution models is the Gaussian Dispersion Plume Model (GDPM).  This note focuses on the development of a deeper understanding of this venerable yet highly robust model via the Maximum Entropy Principle.  The GDPM is a mathematical model that utilizes a Gaussian distribution to describe temporal and spatial turbulent diffusive and advective concentration spread of the pollutants from a number of sources constrained by initial and boundary conditions..  This model has been applied to a wide number of situations such as the release of smoke from wildfires \cite{atmos9050166,Lee2019}, carbon dioxide from volcanoes \cite{Jiang2019}, radionuclides from the Fukushima and Chernobyl nuclear accidents \cite{KORSAKISSOK2013267,Shamsuddin2017,OMARNAZIR201855}, chemical or biological agents\cite{Ma2017, Wein4346,Ray2011,Hiromassa2009,Creary2019} in the atmosphere in response to various meteorological conditions. These models are widely used by numerous agencies such as the United States' Environmental Protection Agency (EPA) \cite{EPA}, National Oceanic and Atmospheric Administration (NOAA) \cite{NOAA1981}, the Department of Defence (DoD) \cite{DoD} as well as numerous private industries.  

\paragraph{Outline}
The GDPM is based on the Gaussian distribution and as such will be discussed first.  The mathematics of the model will be examined in detail. The Maximum Entropy concept will be introduced as well a generalized analytic solution derived from the paradigm.  This maximum entropy solution will then be applied to various transport phenomena such as the advection and advection-diffusion equations in one and several dimensions.  The results from these applications will then be applied to the full GDPM which will be derived from first principles.  The results will be analysed along with experimental techniques to refine the model.

\section{The Importance of Being Gaussian}
The Gaussian distribution, also known as the normal distribution, the central distribution, and the bell curve due to its familiar shape resembling that of a bell, is a pervasive probability density distribution (pdf), $p(x)$ that yields the expected frequency of an outcome as a function of some variable $x$.  This distribution can be found in a myriad of subjects such as physics, economics, biology, mathematics, statistics etc.  This ever-present distribution is found nearly everywhere for a number of reasons that will be examined below.  

\begin{figure}
\centering
\includegraphics[scale=0.35]{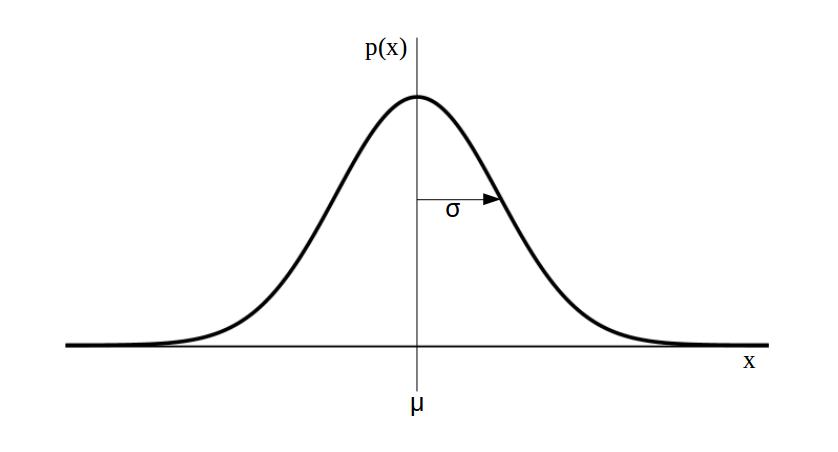}
\caption{An example of a Gaussian distribution.  The center of the distribution is given by the average $\mu$ and the width of the distribution is given by the square root of the variance (standard deviation) $\sigma$.}
\label{fig:gaussian}
\end{figure}
The Gaussian distribution is symmetric and resembles that of the bell curve and is often described by this phrase.  Mathematically the Gaussian distribution, in one-dimension, is described by
\begin{equation}
p(x) = \frac{1}{\sqrt{2\pi\sigma^{2}}}e^{-\frac{1}{2}\frac{(x-\mu)^{2}}{\sigma^{2}}},
\end{equation}
where $\sigma$ is the standard deviation, $\mu$ is the mean, and $x$ is the independent variable corresponding to the probability of $p(x)$ (see Figure \ref{fig:gaussian}).  The Gaussian distribution has a number of mathematically attractive traits such as:
\begin{itemize}
 \item the mean, median, and mode are the same,
 \item the distribution is symmetric about the average $x=\mu$,
 \item the distribution is completely described by two moments,
 \item even ordered higher moments are related to standard deviation and mean,
 \item the product of Gaussian distributions is again a Gaussian distribution,
 \item the convolution of Gaussian distributions is again a Gaussian distribution,
 \item the Fourier transform of a Gaussian is again a Gaussian,
 \item the sum of Gaussian distributions is again a Gaussian distribution.
\end{itemize}

The Gaussian distribution is a result of the central limit theorem.  In essence, this theorem states that as the number of trial samples becomes large, the probability distribution approaches that of the Gaussian distribution.  In many situations there are a number of independent random variables that follow some distribution law themselves.  Each of these independent trials are sampled from their own distribution and then combine to form the overall Gaussian distribution.  This could be thoughts of intuitively as the sum of Gaussian distributions simply yield another Gaussian. 
The Gaussian, $y = e^{-ax^{2}}$, also appears as a solution to differential equations of the form
\begin{equation}
\frac{dy}{dx}+2axy=0.
\end{equation}
This class of differential equations occurs when the rate of change of some quantity can be related back to itself.  Examples of this kind of relationship can be found in radioactivity, material cooling, population dynamics, and interest from bank accounts.

Due to the robustness of the Gaussian distribution, even distributions that are not Gaussian can many times be approximated by this distribution.  Discrete distributions like the binomial and Poisson distributions will asymptotically approach the Gaussian distribution, as will continuous distributions like log-normal, Rayleigh, Gamma, Laplace, chi-square, and student t- distributions. Again, this is due to the central limit theorem described above.  Given a sufficiently large number of random events with some variance and mean, the result will become Gaussian distributed.

\section{The Mathematical Model} \label{mathmodel}
Pollution in the atmosphere is essentially described by a conservation law for the suspended particles in the environmental media.  This combination of conservation of mass and 
the advection-diffusion equation describing turbulent diffusion can be written as,
\begin{equation}\label{eqn:generic_adv_diff}
\frac{\partial C}{\partial t} + \vec{\nabla}\cdot(\vec{u}C) = \vec{\nabla}\cdot(\vec{K}\vec{\nabla}C),
\end{equation}
where $C$ is the concentration, $\vec u =(u,v,w)$ is the average wind velocity with components in the x-,y-, and z-directions respectively, and $K =(K_x,K_y,K_z)$ is the eddy diffusivity with components in the x-,y-, and z-directions respectively.  Assuming the medium is incompressible and the divergence operator can be expanded such that $\vec{\nabla}\cdot\vec{u}=0$, Eqn \ref{eqn:generic_adv_diff} can be written as,
\begin{equation} \label{advection_diffusion_eqn}
\frac{\partial C}{\partial t} + u\frac{\partial C}{\partial x} + v\frac{\partial C}{\partial y} + w\frac{\partial C}{\partial z}= \frac{\partial}{\partial x}\Big(K_{x}\frac{\partial C}{\partial x}\Big) + \frac{\partial}{\partial y}\Big(K_{y}\frac{\partial C}{\partial y}\Big) + \frac{\partial}{\partial z}\Big(K_{z}\frac{\partial C}{\partial z}\Big).
\end{equation}
It should be noted that the wind velocity and concentration are made up of two components,
\begin{equation}
    \vec{u} = \vec{u}_{avg} + \vec{u}_{fluct}
\end{equation}
and 
\begin{equation}
    C = C_{avg} + C_{fluct},
\end{equation}
associated with the average (avg) and fluctuating (fluct) components of the parameter being examined.  It is assumed that the fluctuations cancel out on large enough time scales and   ignore their contribution.

The bulk motion in the $x$-direction dominates over the contribution from the corresponding diffusion term in the same direction and so the diffusion term is neglected.  It is assumed that the wind speeds in the $y-$  and $z-$ directions are minimal in comparison to the wind speed in $x$ and so they are neglected. This yields, 
\begin{equation} \label{advection_diffusion_eqn2}
\frac{\partial C}{\partial t} + u\frac{\partial C}{\partial x} =  \frac{\partial}{\partial y}\Big(K_{y}\frac{\partial C}{\partial y}\Big) + \frac{\partial}{\partial z}\Big(K_{z}\frac{\partial C}{\partial z}\Big).
\end{equation}
where the $x$-direction is the direction of wind current (along-wind), the $y$-direction is the cross-wind direction, the $z-$direction is the vertical direction measured from the ground, $C$ is the concentration of the pollutant, $K_y$ and $K_z$ are the diffusivity in the $x$ and $y$-directions, and $u$ is the mean wind velocity along the $x-$axis.   Assuming that the diffusivity constants are indeed constant leads to the equations
\begin{equation} \label{advection_diffusion_eqn2}
\frac{\partial C}{\partial t} + u\frac{\partial C}{\partial x} =  K_{y}\frac{\partial^2 C}{\partial y^2} + K_{z}\frac{\partial^2 C}{\partial z^2}.
\end{equation}
The boundary conditions are: 
\begin{itemize}
    \item the source concentration is given as $C(0,y,z)=\frac{Q}{u}\delta(y)\delta(z-H)$ where Q is the concentration rate from a source located at a height $z=H$ at position $x=0$ and advected at a velocity $u$ in the $x-$direction,
    \item $C(x,\pm \infty,z)=0$,
    \item $C(\infty,y,z)=0$
    \item $C(x,y,\infty)=0$,
    \item and $K_{z}\frac{\partial C}{\partial z}(x,y,0)=0$ which is a statement that the ground is reflecting.
\end{itemize}
This yields the celebrated solution along with the boundary conditions, defining the standard deviation, $\sigma$, in terms of the diffusivity constant and time as $\sigma^{2}=2Kt$,
\begin{equation} \label{gdpm_soln}
C(x,y,z) = \frac{Q}{2\pi u \sigma_y \sigma_x}\left[e^{-\frac{y^{2}}{2\sigma_{y}^{2}}}\right]\left(e^{-\frac{(z-H)^2}{2\sigma_{z}^{2}}}+e^{-\frac{(z+H)^2}{2\sigma_{z}^{2}}}\right),
\end{equation}
where $x$ is the downwind distance, $y$ is the crosswind distance, $z$ is the vertical distance from the ground, $Q$ is the contaminant emission rate, $\sigma_{y}$ is the lateral dispersion function, $\sigma_{z}$ is the vertical dispersion function, and $H$ is the effective stack height. 

It should be noted that the effective stack height is made up of two contributions, 
\begin{equation}
    H= h+\delta h_{pr}
\end{equation}
where $h$ is the actual stack height and $\delta h_{pr}$ is the effective plume rise above the top of the smoke stack.

It should also be noted that Eq. \ref{gdpm_soln} reduces to the solution for an emission point source at ground level without an effective plume rise,
\begin{equation}
    C(x,y,z) = \frac{Q}{\pi u \sigma_y \sigma_x}e^{-\frac{1}{2}\left(\frac{y^{2}}{\sigma_{y}^{2}}+\frac{z^{2}}{\sigma_{z}^{2}}\right)}
\end{equation}
where the parameters are defined as above.

\section{The Maximum Entropy Principle}
Entropy may simply be regarded as an information measure for  a system.
The MaxEnt principle finds its roots in Laplace's principle of insufficient reason (or sometimes referred to as indifference) which states that, given no additional information, it is reasonable to choose a uniform probability distribution.  The MaxEnt principle takes this to its natural conclusion and suggests that the probability distribution is the least biased distribution given the known constraints.  

Jaynes \cite{Jaynes1957_1, Jaynes1957_2} suggested that the most likely probability distribution function should be one that obeys the known constraints (the moments) while maximizing the entropy.  This is known as the maximum entropy (MaxEnt) method.  This technique minimizes the amount of relevant prior information that is used to determine the distribution function while allowing the distribution to evolve towards maximal entropy.  This idea of finding the least biased distribution with the known data as constraints is highly appealing.  

It should be noted that Maximum Entropy techniques have been applied to a wide variety of subjects.  In physics, this technique has been used in particle physics to determine the quark distribution in pions \cite{HAN2020}, QCD calculations \cite{DING2019}, in condensed matter physics applications such as solving certain mathematical techniques \cite{LEVY2017} and the modeling of Helium-4 \cite{PhysRevB.kora}and in astronomy the principle has been used to aid in the reconstruction of x-ray images \cite{MRAS_Willingale, Guan_2016} and in the analysis of the mass-radius relationship of exo-planets \cite{Ma_2019},  as well as providing a unified view of transport equations( ref our paper). In biology it has been used to analyse the structure of DNA \cite{yeoburge2004,trebeau2019,Levy2019}, model species distributions \cite{Levy2019, wang2018,Kalboussi2018,Baldwin2009}, and to study gene expression \cite{Zhu2010,RODRIGUEZ2019}.   The method has even been applied to understanding economic growth in a myriad of settings \cite{ALAM2017, Chakpitak_2017, YALTA2011}. 

The constraints, which represent the limited amount of information known, may be the moments of the probability distribution function\cite{SECREST2020}. These constraints can be experimentally determined.  Analytically, the $n^{th}$ moment, $\mu_{n}$, depending on a physical observable $\xi$ for a probability distribution function $\rho(\xi)$ is defined as
\begin{equation}
\label{eqn:moments}
\langle \xi^{n} \rangle = \int_{-\infty}^{\infty} \xi^{n}\rho(\xi)d\xi.
\end{equation}
It should be noted that the zeroth moment, $\langle 1 \rangle$ is the normalization.
The Shannon-Kullback information entropy is defined as \cite{Shannon1948} the   weighted average of the log of the probabilities,
\begin{equation}
   S = -\int_{-\infty}^{\infty}\rho(\xi) \text{ln} \frac{\rho(\xi)}{\rho_{0}(\xi)} d\xi,
\end{equation}
 where $\rho_{0}$ is an invariant measure.  The maximum entropy principle states that the most likely probability distribution will be one that maximizes the entropy subject to  the known constraints.  The constraints are imposed via Lagrange's multipliers(which may have time dependence) $\lambda_{0}, \lambda_{1},$ and $\lambda_{2}$, 
\begin{equation}
\label{eqn:S}
\begin{aligned}
S = &-\int_{-\infty}^{\infty}\rho(\xi) \text{ln} \frac{\rho(\xi)}{\rho_{0}(\xi)} d\xi  \\
&-(\lambda_{0}-1)\left[\int_{-\infty}^{\infty}\rho(\xi)d\xi - \langle 1 \rangle \right] \\
& -\lambda_{1}\left[\int_{-\infty}^{\infty}\xi \rho(\xi)d\xi - \langle \xi^{1} \rangle \right] \\
& -\lambda_{2}\left[\int_{-\infty}^{\infty}\xi^{2}\rho(\xi)d\xi - \langle \xi^{2} \rangle \right].
\end{aligned}
\end{equation}
Variational calculus is used in order to maximize Eq. (\ref{eqn:S}) within the bounds of the constraints,
\begin{equation}
\label{eqn:max_S}
\delta S = 0,
\end{equation}
and solving for the probability distribution yields
\begin{equation}
\label{eqn:MaxEnt_soln}
\rho = \rho_{0}(\xi)e^{-\lambda_{2}\xi^{2} - \lambda_{1}\xi - \lambda_{0}}.
\end{equation}
This yields a generic solution to the MaxEnt problem.  One must determine the Lagrange multipliers from the equations of constraint  in order to determine the particular solution at hand.

If the time derivatives of the equations of constraint  constitute a closed set of differential equations  the static solution  may be viewed as the solution to the MaxEnt equations at a given time and the time dependent solutions are obtained by replacing $\lambda_i$ by $\lambda_i(t)$ which  satisfy the time dependent equations of constraint (\cite{AandO:2001}).

\section{Application of Maximum Entropy Principle to One-Dimensional Advection}\label{sec:one_d_advection}
The advection equation \cite{bennett2012transport} is a hyperbolic partial differential equation of the form
\begin{equation}
\label{eqn:advection_eqn}
\frac{\partial \rho(x,t)}{\partial t} + v \frac{\partial \rho(x,t)}{\partial x} =0
\end{equation}
that describes how a scalar field density $f$ is swept along (advected) by a 
bulk flow of constant speed $v$.  The position $-\infty < x < \infty$, the time 
$0 < t < \infty$, and the velocity field $v$ are nonzero. Examples of where the 
advection equation is used are modeling automobile traffic\cite{jouelai_2014,champagne_2010}, blood flow through 
a capillary \cite{schelin_2009,bagchi_2007}, and salinity propagation in the ocean \cite{wadley_2006,levitus_1986}. 

It is well known that the exact form of the solution to this partial differential equation has the form $\rho(x,t) = \rho(x-vt,0)$.  This means hat the probability distribution function does not change its initial shape during advection.

Lets apply the MaxEnt method to solve the advection equation, Eq. \ref{eqn:advection_eqn}, to determine the particular solution,
\begin{equation}
\label{eqn:adv_soln}
\rho(x,t) =\frac{1}{\sqrt{2\pi\sigma^{2}}} e^{-\frac{1}{2}\frac{(x-vt)^2}{\sigma^{2}}}.
\end{equation}
given the initial condition of a Gaussian profile $\rho_{0}(x,0)=\frac{1}{\sqrt{2\pi\sigma^{2}}}e^{-\frac{1}{2}x^2}$.

The advection equation is rearranged,
\begin{equation}
\label{eqn:advection_eqn_rearraged}
\frac{\partial \rho(x,t)}{\partial t} = -v \frac{\partial \rho(x,t)}{\partial x} 
\end{equation}
and integrated spatially over $x$ on both sides of the equation,
\begin{equation}
\int_{-\infty}^{\infty}\frac{\partial \rho(x,t)}{\partial t}dx = -\int_{-\infty}^{\infty}v \frac{\partial \rho(x,t)}{\partial x}dx.
\end{equation}
Assuming that the derivative operator can be removed from under the spatial integral along with the constant velocity,
\begin{equation}
\frac{d}{dt}\int_{-\infty}^{\infty} \rho(x,t)dx = -v \int_{-\infty}^{\infty}\frac{\partial \rho(x,t)}{\partial x}dx,
\end{equation}
it is noted that the integral on the left is the zeroth moment, the normalization, and that the integral on the right yields zero due to the boundary conditions imposed on the probability distribution function.  This simply yields,
\begin{equation}
\frac{d\langle 1 \rangle}{dt}=0
\end{equation}
which is just the statement that the zeroth moment, i.e. the normalization, is a constant (usually taken to be one),
\begin{equation}
\label{eqn:adv_m_0}
\langle 1 \rangle=1.
\end{equation}

Again, applying the same idea to Eq.(\ref{eqn:advection_eqn_rearraged}), but this time multiplying both sides by $x$ and integrating,
\begin{equation}
\int_{-\infty}^{\infty}\frac{\partial \rho(x,t)}{\partial t}xdx = -\int_{-\infty}^{\infty}v \frac{\partial \rho(x,t)}{\partial x}xdx,
\end{equation}
assuming that the temporal derivative operator can be removed from the integral on the left and integrating-by-parts on the right yields,
\begin{equation}
\frac{d}{dt}\int_{-\infty}^{\infty}\rho x dx = -v\Big[x\rho|_{-\infty}^{\infty} -\int_{-\infty}^{\infty}\rho dx\Big],
\end{equation}
where the surface goes to zero due to the boundary conditions imposed on the probability distribution function, $\rho$, the integral on the right hand side is identified with the first spatial moment, $\langle x \rangle$, and the remaining integral on the right is identified with the zeroth moment, the normalization $\langle 1 \rangle$.  Note that the differential equations for the moments are a closed set. Finally this results in 
\begin{equation}
\label{eqn:adv_first_int}
\frac{d\langle x \rangle}{dt} = v.
\end{equation}
This is integrated to determine the temporal dependence of the first moment,
\begin{equation}
\label{eqn:adv_m_2}
\langle x \rangle = vt.
\end{equation}

The moments are calculated from the MaxEnt solution are:
\begin{eqnarray}
\label{eqn:m_0}
 \langle 1 \rangle &=& \int_{-\infty}^{\infty}\rho_{0}(x)e^{-\lambda_{2}x^{2} - \lambda_{1}x - \lambda_{0}}= \int_{-\infty}^{\infty} \frac{1}{\sqrt{2\pi\sigma^{2}}}e^{-\frac{1}{2}x^2} e^{- \lambda_{1}x - \lambda_{0}}=e^{-\lambda_{0}}e^{\frac{\lambda_{1}^{2}\sigma^{2}}{2}}, \\
 \text{and}\nonumber \\
\label{eqn:m_2}
\langle x \rangle &=& \int_{-\infty}^{\infty}x\rho_{0}(x)e^{-\lambda_{2}x^{2} - \lambda_{1}x -\lambda_{0}}= \int_{-\infty}^{\infty} x\frac{1}{\sqrt{2\pi\sigma^{2}}}e^{-\frac{1}{2}x^2} e^{- \lambda_{1}x - \lambda_{0}}=-\lambda_{1}\sigma^{2}e^{-\lambda_{0}}e^{\frac{\lambda_{1}^{2}\sigma^{2}}{2}}.
\end{eqnarray}

Equating the moments from Eqs. (\ref{eqn:adv_m_0}) and (\ref{eqn:m_0}) and Eqs. (\ref{eqn:adv_m_2}) and (\ref{eqn:m_2}) and solving for the Lagrange multipliers, $\lambda_0$=
$\lambda_{0}(t)=\frac{(vt)^{2}}{2\sigma^{2}}$ and $\lambda_1$=$\lambda_{1}(t)=-\frac{vt}{\sigma^{2}}$.  Substituting these results back into the MaxEnt solution Eq. (\ref{eqn:MaxEnt_soln}) results in the 
solution given by Eq. (\ref{eqn:adv_soln}).

\section{Application of Maximum Entropy Principle to One-Dimensional Diffusion}\label{sec:one_d_diffusion}
Consider the one-dimensional diffusion equation with constant eddy diffusivity constant $K$ is given by
\begin{equation}
\label{eqn:diff_eqn}
\frac{\partial \rho}{\partial t} = K\frac{\partial^{2}\rho}{\partial x^{2}},
\end{equation}
where $\rho$ is the probability distribution function that obeys the boundary conditions that as $x$ approaches $\pm\infty$, $\rho \rightarrow 0$ with the initial condition that $\rho(x,t) = \rho(x,0)=\rho_{0}$. This results in the celebrated Gaussian solution,
\begin{equation}
\label{eqn:diff_soln}
    \rho=\frac{\rho_{0}}{\sqrt{2\pi}\sigma} e^{{-\frac{1}{2}}\Big(\frac{x}{\sigma}\Big)^{2}}.
\end{equation}

Integrating both sides of Eq. (\ref{eqn:diff_eqn}),
\begin{equation}
\int^{\infty}_{-\infty}\frac{\partial \rho}{\partial t} dx = K\int^{\infty}_{-\infty}\frac{\partial^{2}\rho}{\partial x^{2}}dx,
\end{equation}
pulling out the temporal derivative operator and noting the integral on the left hand side is the zeroth moment of the probability distribution function, $\int^{\infty}_{-\infty}\rho dx = \langle 1 \rangle$,
\begin{equation}
\frac{d \langle 1 \rangle }{dt}= \frac{\partial \rho}{\partial x} \Big|^{\infty}_{-\infty} = 0
\end{equation}
simply yields the normalization condition,
\begin{equation}
\langle 1 \rangle = 1.
\end{equation}
Multiplying both sizes by the position $x$, integrating in order to exploit the first moment,
\begin{equation}
\frac{d }{dt}\int^{\infty}_{-\infty}x \rho dx = K\int^{\infty}_{-\infty}x\frac{\partial^{2}\rho}{\partial x^{2}}dx
\end{equation},
noting that the integral on the left is the average position, $\langle x \rangle$, and integrating by parts on the right hand side gives
\begin{equation}
\frac{d \langle x \rangle }{dt}= K\left[x\frac{\partial \rho}{\partial x}\Big|^{\infty}_{-\infty} - \int^{\infty}_{-\infty}\frac{\partial \rho}{\partial x}dx \right]=0
\end{equation}
It is seen that the average position in the $x$-direction is a constant.  This is to be expected for a symmetric distribution starting at $x=0$.\\

Applying this technique again to Eq. (\ref{eqn:diff_eqn}) by multiplying both sides by $x^{2}$ and integrating over all space to exploit the second position moment gives
\begin{equation}
\frac{d}{dt}\int^{\infty}_{-\infty}x^{2}\rho dx = K\int^{\infty}_{-\infty}x^{2}\frac{\partial^{2}\rho}{\partial x^{2}}dx.
\end{equation}
Again noting that the left hand side is the second moment and using integration-by-parts on the right hand side results in
\begin{equation}
\frac{d \langle x^{2} \rangle}{dt} = K\left[x^{2}\frac{\partial \rho}{\partial x}\Big|^{\infty}_{-\infty} - \int^{\infty}_{-\infty}2x\frac{\partial C}{\partial x}dx \right]
\end{equation}
and integrating the second integral by parts yields
\begin{align}
\frac{d \langle x^{2} \rangle}{dt}  & =  K\Big[ -2\Big(\rho x\Big|^{\infty}_{-\infty}-\int^{\infty}_{-\infty}\rho dx\Big)\Big]\\
                                    & = 2K\langle \rho \rangle.
\end{align}
Integrating, applying the initial conditions, and solving for the second moment results in 
\begin{equation}
\langle x^{2} \rangle = 2Kt.
\end{equation}
Knowing that the solution from Eq. (\ref{eqn:MaxEnt_soln}) has the form
\begin{equation}
\rho = \rho_{0}e^{-\lambda_{2}x^{2} - \lambda_{0}},
\end{equation}
for the initial condition $\rho_{0}$,
one can use the constraints and the generic solution to determine the undetermined multipliers. From the normalization
\begin{equation}
\langle 1 \rangle = \int^{\infty}_{-\infty} \rho dx =  \int^{\infty}_{-\infty}\rho_{0}e^{-\lambda_{2}x^{2} - \lambda_{0}} =\rho_{0}e^{-\lambda_{0}}\left[\frac{\sqrt{\pi}}{\sqrt{\lambda_{2}}}\right]=1
\end{equation}
and the second moment along with the result
\begin{equation}
\langle x^{2} \rangle = \int^{\infty}_{-\infty}x^{2}\rho dx =  \rho_{0}e^{- \lambda_{0}}\int^{\infty}_{-\infty}x^{2}e^{-\lambda_{2}x^{2}}dx=\rho_{0}e^{- \lambda_{0}}\left[\frac{\sqrt{\pi}}{2\lambda_{2}^{\frac{5}{2}}}\right]=2\rho_{0}Kt
\end{equation}
one finds $e^{-\lambda_{0}}=\frac{\sqrt{\lambda_{2}}}{\sqrt{\pi}}$ and $\lambda_{2}=\frac{1}{4Kt}$, which yields the well-known solution
\begin{equation}
\rho =\frac{\rho_{0}}{\sqrt{4\pi Kt}}e^{-\frac{x^{2}}{4Kt}}.
\end{equation}
Typically this is in terms of the standard deviation $\sigma=\sqrt{2Kt}$, which is the solution given by Eq. (\ref{eqn:diff_soln}). 

In the above example, the diffusive process is centered around the origin, $x=0$.  The probability distribution can be translated to a different center, $x_{0}$ by replacing $x$ with $x-x_{0}$, which amounts to replacing Eq. (\ref{eqn:diff_soln}) with
\begin{equation}\label{eqn:diffusion_result}
    \rho = \frac{\rho_{0}}{\sqrt{2\pi}\sigma}e^{-\frac{(x-x_{0})^{2}}{2\sigma^{2}}}.
\end{equation}

Note that the solutions to the advection equation and the diffusion equation look remarkably similar.  The results of the advection equation were with the initial condition that the probability distribution function was a Gaussian distribution.  Then as time moved forward the solution is simply the same shape of the initial distribution but translated (or better said, advected)  along the $x$-axis.  The solution for the diffusion equation is the Gaussian distribution centered around its initial position (in this case $x=0$), an initial condition was not given and it should be noted that the distribution widens as time increases. Hence the solutions may appear the same at first but the origin of the two solutions is different, as are the subsequent characteristics of each.

\section{Application of Maximum Entropy Principle to One-Dimensional Advection-Diffusion Model}\label{sec:one_d_advection_and_diffusion}
\begin{figure}[h]

\centering
\includegraphics[scale=0.25]{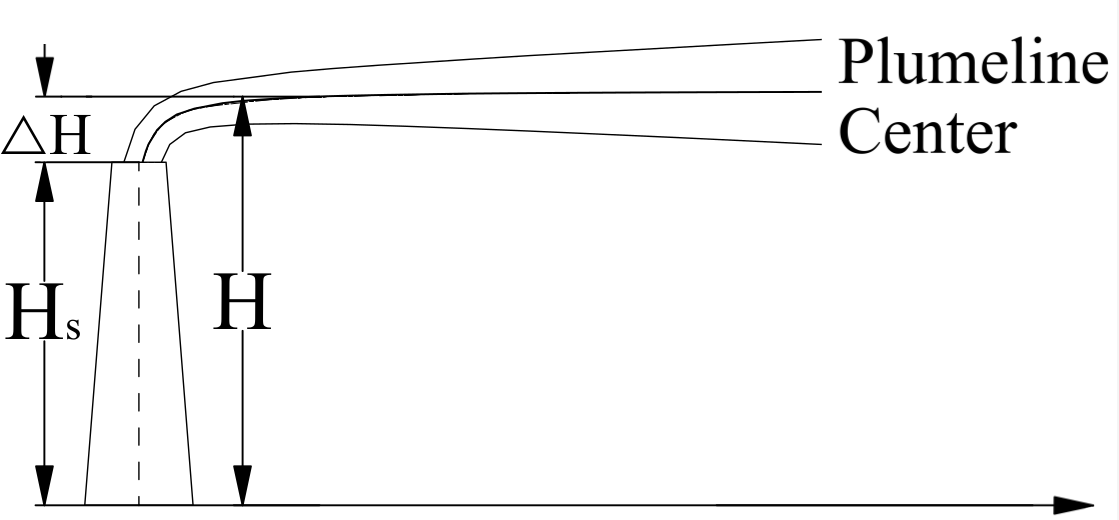}
\caption{The point source for emission is given by the height of the smoke stack.  The plume is advected horizontally along the plumeline center as the width of the smokeplume grows due to diffusion.}
\label{fig:1d_setup}
\end{figure}

The one-dimensional advection-diffusion equation is given by
\begin{equation}
\label{eqn:one_dim_adv_diff}
\frac{\partial \rho}{\partial t} + v\frac{\partial \rho}{\partial x} = K \frac{\partial^{2} \rho}{\partial x^{2}},
\end{equation}
where $\rho$ is the probability density function, $v$ is the advective velocity, and $K$ is the eddy diffusion constant (see Figure \ref{fig:1d_setup}).
Integrating both sides of Eq. (\ref{eqn:one_dim_adv_diff}) by $x$,
\begin{equation}
\frac{d}{dt}\int^{\infty}_{-\infty}  \rho dx + v\int^{\infty}_{-\infty} \frac{\partial \rho}{\partial x}dx = K \int^{\infty}_{-\infty} \frac{\partial^{2} \rho}{\partial x^{2}} dx,
\end{equation}
yields the result
\begin{equation} \label{eqn:int_zero_moment}
\frac{d}{dt}  \langle 1 \rangle =  0,
\end{equation}
implying that the normalization is a constant.  Again multiplying by $x$ and integrating Eq. (\ref{eqn:one_dim_adv_diff}) gives
\begin{equation}
\frac{d}{dt}\int^{\infty}_{-\infty} x \rho dx + v \int^{\infty}_{-\infty} x\frac{\partial \rho}{\partial x}dx = K \int^{\infty}_{-\infty} x \frac{\partial^{2} \rho}{\partial x^{2}} dx.
\end{equation}
Integrating by parts leads to
\begin{equation}
\frac{d}{\partial t} \langle x \rangle +\left[x\rho\Big|^{\infty}_{-\infty}-\int^{\infty}_{-\infty} \rho dx\right] = K\left[x\frac{\partial \rho}{\partial x}\Big|^{\infty}_{-\infty} - \int^{\infty}_{-\infty}\frac{\partial \rho}{\partial x} dx\right],
\end{equation}
where the surface terms go to zero due to the boundary conditions.  After integrating over time, this leads to 
\begin{equation}
\label{eqn:int_first_moment}
<x> = vt.
\end{equation}
Multiplying Eq. (\ref{eqn:one_dim_adv_diff})  by $x^{2}$ to obtain an equation in terms of the second moment and integrating leads to
\begin{equation}
\frac{d}{dt}\int^{\infty}_{-\infty} x^{2} \rho dx + u\int^{\infty}_{-\infty} x^{2}\frac{\partial \rho}{\partial x}dx = K \int^{\infty}_{-\infty} x^{2} \frac{\partial^{2} \rho}{\partial x^{2}} dx.
\end{equation}
Again, integrating by parts yields
\begin{equation}
\frac{d }{dt} \langle x^2 \rangle +\left[x^{2}\rho\Big|^{\infty}_{-\infty}-2\int^{\infty}_{-\infty} x\rho dx\right] = K\left[x^{2}\frac{\partial \rho}{\partial x}\Big|^{\infty}_{-\infty} - 2\int^{\infty}_{-\infty} x\frac{\partial \rho}{\partial x} dx\right].
\end{equation}
Again, the surface terms go to zero. Applying integration by parts to the last integral on the right hand side leads to
\begin{equation}
\frac{d}{dt}\langle x^{2} \rangle -2v\langle x\rangle = K\left[-2\left(x \rho\Big|^{\infty}_{-\infty}-\int^{\infty}_{-\infty} \rho dx \right) \right].
\end{equation} 
Using the result from Eq. (\ref{eqn:int_first_moment}) and integrating over time yields
\begin{equation}\label{eqn:int_second_moment}
\langle x^{2} \rangle = 2 Kt + v^{2}t^{2}.
\end{equation}
Using the MaxEnt solution of Eq. (\ref{eqn:MaxEnt_soln}) to find the zeroth, first, and second moments, and equating to Eqs. (\ref{eqn:int_zero_moment}), (\ref{eqn:int_first_moment}), and (\ref{eqn:int_second_moment}), results in
\begin{equation}
\langle 1 \rangle = \int^{\infty}_{-\infty} \rho dx =  \int^{\infty}_{-\infty}\rho_{0}e^{-\lambda_{2}x^{2} -\lambda_{1}x- \lambda_{0}} =\rho_{0}e^{-\lambda_{0}}\left[\frac{e^{\frac{\lambda^{2}_{1}}{4\lambda_{2}}}\sqrt{\pi}}{\sqrt{\lambda_{2}}}\right]=1,
\end{equation}
\begin{equation}
\langle x \rangle = \int^{\infty}_{-\infty}x^{2}\rho dx =  \rho_{0}\int^{\infty}_{-\infty}xe^{-\lambda_{2}x^{2}-\lambda_{1}x-\lambda_{0}}dx=-\rho_{0}e^{- \lambda_{0}}\left[\frac{\lambda_{1}e^{\frac{\lambda^{2}_{1}}{4\lambda_{2}}}\sqrt{\pi}}{4\lambda_{2}^{\frac{3}{2}}}\right]=vt,\\
\end{equation}
and
\begin{equation}
\langle x^{2} \rangle = \int^{\infty}_{-\infty}x^{2}\rho dx =  \rho_{0}\int^{\infty}_{-\infty}x^{2}e^{-\lambda_{2}x^{2}-\lambda_{1}x-\lambda_{0}}dx=\rho_{0}e^{- \lambda_{0}}\left[\left(2\lambda_{2}+\lambda_{1}^{2}\right)\frac{e^{\frac{\lambda^{2}_{1}}{\lambda_{2}}}\sqrt{\pi}}{4\lambda_{2}^{\frac{5}{2}}}\right]=2Kt+v^{2}t^{2}\\.
\end{equation}
Solving for the Lagrange multipliers leads to $e^{-\lambda_{0}}=\frac{1}{\sqrt{4\pi Kt}}e^{-\frac{v^2t^2}{Kt}}$, $\lambda_{1}=\frac{v}{2K}$, and $\lambda_{2}=\frac{1}{4Kt}$.  Finally, this leads to the result that the probability distribution function is
\begin{equation}
\frac{\rho_{0}}{\sqrt{4\pi Kt}}e^{-\frac{(x-vt)^{2}}{4Kt}},
\end{equation}
which is the combination of the advection and diffusion equation solutions Eqs. (\ref{eqn:adv_soln}) and (\ref{eqn:diff_soln}) in one dimension that were solved in Sections \ref{sec:one_d_advection} and \ref{sec:one_d_diffusion}.

\section{Application of Maximum Entropy Principle Results to the Multi-Dimensional Advection-Diffusion Model}
\begin{figure}
\centering
\includegraphics[scale=0.25]{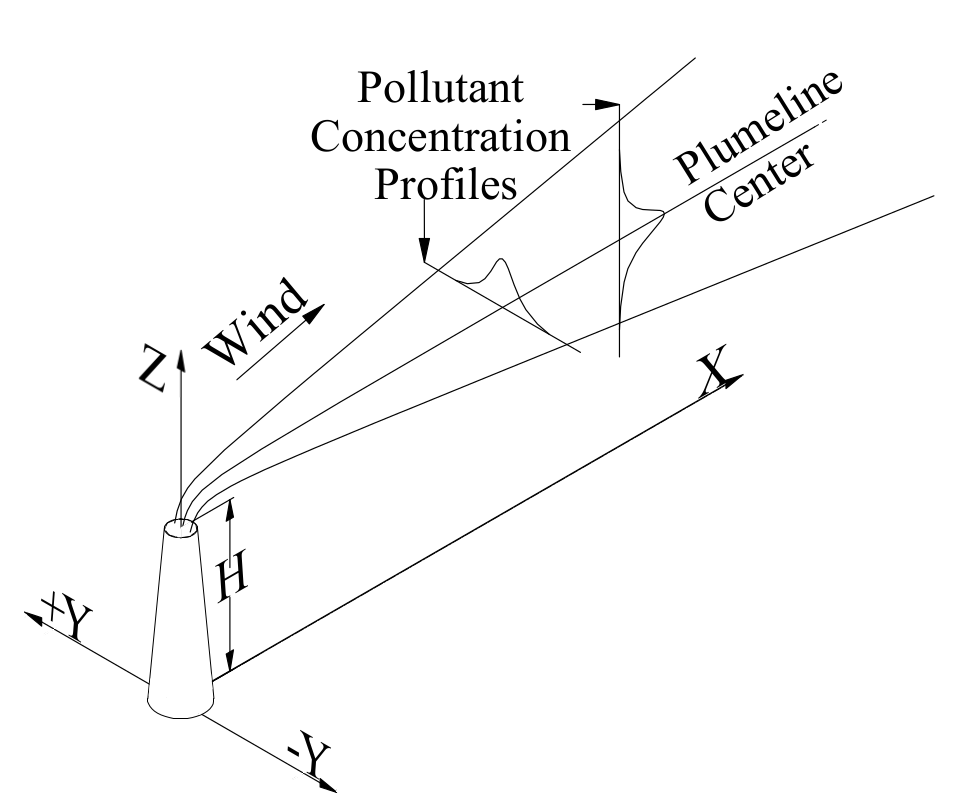}
\caption{The smoke plume source along with the concentration profile in the coordinate system under consideration.}
\label{fig:3d_source}
\end{figure}

Using the MaxEnt results from Sections \ref{sec:one_d_advection}, \ref{sec:one_d_diffusion}, and \ref{sec:one_d_advection_and_diffusion}, the three dimensional GDPM can be derived.  See Fig. \ref{fig:3d_source} for the configuration of the physical system under consideration.

\subsection{Source}
The strength of the initial emission $Q$ can be handled in one of two different ways in the model.  One method is to add a source term, $S$ and use Dirac delta functions of the form $S = Q\delta(x-x_{0})\delta(y-y_{0})\delta(z-z_{0})$ to Eq. (\ref{advection_diffusion_eqn}) .  Another method which will be employed in this work will be to use the invariant measure in the MaxEnt solution in Eq. (\ref{eqn:MaxEnt_soln}).  In this case the invariant measure is $\rho = \frac{Q}{u}$ where again, Q is the strength of the emission and $u$ is the downwind speed.

\subsection{Reflection from the Ground}\label{reflection_ground}
The Gaussian plume that is emitted from some finite height above the ground will eventually spread and encounter the barrier of the ground as the it moves downwind.  Since the amount of material must be conserved, the effect of the interaction with the contaminant plume with the ground will cause a reflection effect.  Typically in these models the ground is considered a perfect reflector. This ground reflection effect is modeled using the Method of Images \cite{jackson_classical_1999}.  The Method of Images models the reflection by using a virtual source of the strength located the same distance away but below the surface.

\subsection{MaxEnt Solution for the Gaussian Plume Model}
The total concentration of pollutant at a given time and location is given by the product of the concentrations in the three directions:
\begin{equation}
    C(x,y,z,t)=c_{x}(x,t)c_{y}(y,t)c_{z}(z,t)
\end{equation}
where $c_{x}(x,t)$ is the concentration in the x-direction, $c_{y}(y,t)$ is the concentration in the y-direction, and $c_{z}(z,t)$ is the concentration in the z-direction.  The x-component of the concentration is given by the invariant measure that $\rho=\frac{Q}{u}$.  The y-component of the concentration is given by Eq. (\ref{eqn:diff_soln}) in Section \ref{sec:one_d_diffusion}.  The z-component is given by the corresponding version of Eq. (\ref{eqn:diff_soln}) in Section \ref{sec:one_d_diffusion}.  This yields, 
\begin{equation}
\label{gdpm_partial_soln}
C(x,y,z) = \frac{Q}{2\pi u \sigma_y \sigma_x}\left[e^{-\frac{y^{2}}{2\sigma_{y}^{2}}}\right]e^{-\frac{(z-H)^2}{2\sigma_{z}^{2}}}.
\end{equation}
Now applying the Methods of Images and utilizing the law of superposition in adding a virtual source below the ground at $z=-H$, yields the result,
\begin{equation}
\label{gdpm_total_soln}
C(x,y,z) = \frac{Q}{2\pi u \sigma_y \sigma_x}\left[e^{-\frac{y^{2}}{2\sigma_{y}^{2}}}\right]\left(e^{-\frac{(z-H)^2}{2\sigma_{z}^{2}}}+e^{-\frac{(z+H)^2}{2\sigma_{z}^{2}}}\right),
\end{equation}
which is in agreement with well-known result quoted in Eq. (\ref{gdpm_soln}). 

\section{A Novel Way to Apply the Maximum Entropy Principle to the Gaussian Plume Model}
The GDPM is an excellent example of the application of MaxEnt principle where the information is maximized within the constraints of the situation at hand.  These constraints are given by the low-lying moments of the distribution of interest.  Since the MaxEnt solution is known, in general, all that must be done in order to determine the particular solution is to determine the constraints via the moments of the distribution.  This was recently described by the authors (\cite{SECREST2020}) where the partial differential equations of interest did not need to be solved. Instead the MaxEnt solution can be used, along with the experimentally determined low-lying moments, to describe the dynamics of the problem.

The GDPM has a long history of doing just that.  Experiments have been carried out where concentrations of particulates have been measured at regular intervals depending on various variables such as rural or urban environments, day or night, and the stability of the atmosphere (\cite{meade_and_pasquill_1958, Pasquill1961, Gifford1961, Turner_1979}).  Tables of values have been tabulated and often the variance is represented by a power law or other, more sophisticated,  parameterizations.  These phenomenological models of the low-lying moments are then fed back into the solution in order to yield a realistic representation of the plume dynamics.

\section{Conclusions}
The MaxEnt principle was applied to the the advection, diffusion, and the advection-diffusion equations as well as to the advection-diffusion equation along with the appropriate initial and boundary conditions and the GDPM was recovered.  This yields a deeper insight into the understanding of the model and its application to continuous release point sources.

Applying the MaxEnt Principle to the model yields a deeper understanding as to why the GDPM works as it does.  The fundamental notion is that the probability distribution that one is seeking to model is the one that maximizes the entropy within the given set of constraints in this case the low-lying moments which have been tabulated and parameterized.   

The authors do not see any reason that the MaxEnt principle could not be further applied to more complex sources such as line, area, and volume sources and to other models such as the Box Model (\cite{RAGLAND1973,Seogcheol2005}), Gaussian Puff Model (\cite{JUNG2003}), or more complicated numerical models.

A natural extension of this work is to consider different entropic measures.  In this work the Shannon entropy was employed but other entropies such as the Tsallis (\cite{Tsallis1988}) and Kaniadakis (\cite{Kaniadakis2017}) entropy could be investigated  for fractional transport equations (\cite{Wyss1986,Wei_2016, GOMEZ2016}) to describe the diffusive and advective processes.

\end{document}